\documentclass[twocolumn]{aastex61}
\usepackage{natbib}

\newcommand\aastex{AAS\TeX}

\accepted{May 13, 2018}
\submitjournal{ApJ}

\shorttitle{\aastex\ The emission line region of 3c~186}
\shortauthors{Chiaberge et al.}

\begin{document}

\title{The recoiling black hole candidate 3C~186:
  spatially-resolved quasar feedback and further evidence of a blue-shifted broad line region.}

\correspondingauthor{Marco Chiaberge}
\email{marcoc@stsci.edu}

\author[0000-0003-1564-3802]{Marco Chiaberge}
\affil{Space Telescope Science Institute, 3700 San Martin Dr.,
Baltimore, MD, 21218 USA}
\affil{Johns Hopkins University - Center for Astrophysical Sciences,
  3400 N. Charles Street,
  Baltimore, MD 21218, USA}

\author{Grant R. Tremblay}
\affiliation{Harvard-Smithsonian Center for Astrophysics,
60 Garden Street,
Cambridge, MA 02138, USA}

\author{Alessandro Capetti}
\affiliation{INAF-Osservatorio Astrofisico di Torino,
 Via Osservatorio 20,
 10025, Pino Torinese, Italy}


\author{Colin Norman}
\affil{Space Telescope Science Institute, 3700 San Martin Dr.,
Baltimore, MD, 21218 USA}
\affil{Johns Hopkins University - Center for Astrophysical Sciences,
  3400 N. Charles Street,
  Baltimore, MD 21218, USA}



\begin{abstract}

  We present the results of integral field spectroscopy of the gravitational wave (GW)
  recoiling black hole candidate 3C~186. The goal of the observations is to study
  the kinematics of the [OIII]5007 narrow emission line region (NLR) of the quasar,
  and investigate the origin of the velocity offsets originally measured for different
  UV lines. The results show that i)
  the spatial structure of the NLR is complex. The [OIII]5007 line shows significant
  velocity offsets with respect to the systemic redshift of the source. Different
  components at different velocities (-670, -100, + 75 km s$^{-1}$) are produced in different
  regions of the source. ii) we detect both the narrow and the broad components of
  the H$\beta$ line. The narrow component generally follows the kinematics of the
  [OIII] line, while the broad component is significantly blue-shifted. The peak of the broad line
  is near the blue end, or possibly outside of the sensitivity band of the
  instrument, implying a velocity offset of $\gtrsim$1800 km s$^{-1}$. This result is in
  agreement with the interpretation of the QSO as a GW recoiling black hole. The properties
  of the NLR show that the observed outflows are most likely the effect of radiation pressure on the
  (photoionized) gas in the interstellar medium of the host galaxy. 

\end{abstract}




\section{Introduction} \label{sec:intro}

Supermassive black holes (SMBH) may grow through accretion of matter during short-lived
active phases, and/or via mergers. SMBH mergers are expected to occur as a result of major
galaxy mergers, on time-scales that may be as short as 10 Myr,
depending on the properties of the gas in the
merging galaxies, as shown by recent simulations \citep[see][ for a recent review]{mayer17}.
An intimate relationship between galaxy growth and black hole growth also seems to be required to explain tight
relationships such as the M-$\sigma$ correlation, originally discovered by \citet{ferrarese00} and
\citet{gebhardt00}. When galaxies merge, the two central SMBH first lose energy by
dynamical friction, then by three-body interactions with stars that have appropriate angular momentum in the
region of the parameter space \citep[the so-called loss cone,][]{begelman80}. The pair keeps
scattering stars off until it becomes sufficiently tightly bound
that gravitational radiation is the most
efficient mechanism responsible for energy and angular momentum losses. 
If SMBH bound pairs routinely form as a result of galaxy mergers, a stochastic gravitational
wave background formed by the superposition of the low-frequency GW from each of these systems is also expected.
Pulsar timing array (PTA) experiments \citep{ppta13,epta13,nanograv13} and, in the future, space based GW observatories such as LISA \citep{lisa}
are built to be sensitive to such low frequency GW emission. The details of the mechanisms that
pull the two BHs to the distance at which GW
emission becomes substantial are still poorly  understood. A gas-rich environment may significantly help this process. 
Recent work using simulations show that even in gas-poor environments 
SMBH binaries can merge under certain conditions, e.g. if they formed in major galaxy mergers where the final galaxy 
is non-spherical \citep[][and references therein]{khan11,preto11,khan12,bortolas16}.
However, characteristic time-scales to replenish the loss cone may be longer than a Hubble time.
If the loss cone is not replenished quickly enough for the pair to get sufficiently close and
lose energy via gravitational waves, the pair may stall and never merge.
Since the typical distance at which a pair with M$_1$ $\sim$ M$_2 \sim 10^7$ M$_\odot$
starts to efficiently shrink via GW emission is $\sim 1$pc,
this is known as the final parsec problem \citep{milosavljevicmerrit03}.

Recently, stringent upper limits on the GW background radiation
derived with the Parkes PTA were interpreted as being in tension with the current
paradigm of galaxy merger and SMBH pair formation \citep{shannon15}, but alternative scenarios
that may remove such a tension were also proposed \citep{middleton17,rasskazov17}.
The lack of PTA detections may imply that the timescales for the
pairs to merge is either much shorter, or much longer than expected. In the latter case, it could indicate that
the final parsec problem indeed constitutes an issue. 
Detecting GW radiation from
``stalled'' SMBH pairs corresponds to the so-called ``nightmare scenario'' \citep{dvorkin17}.
More broadly, if this prevents SMBH to
merge, our current understanding on the mechanisms for black hole growth might need significant revisions.
It is therefore extremely important to find (either direct, through GW detectors, or indirect)
evidence of SMBH mergers, especially for black holes of very high mass ($10^8 - 10^9 M_\odot$).

One possible way to obtain observational evidence of SMBH mergers using electromagnetic
radiation is to look for runaway (kicked) black holes.
Depending on the properties (mass, spin) of each of the SMBHs in a merging pair, the resulting merged BH may get a kick
and be ejected at velocities that can be in principle as high as the escape velocity of the host galaxy
\citep{merritt04,madauquataert04,komossa12}. 
This process has been extensively studied with numerical simulations
\citep{campanelli07,blecha11,blecha16,healylousto18,gerosa18} but we are still lacking a confirmed example of such a phenomenon.
If the ejected  BH  is active, we
could in principle observe both an offset AGN and velocity shifts between narrow
and broad lines  \citep{loeb07,volonterimadau08}. Such shifts are expected because 
the broad-line emitting region is dragged out with the kicked BH, while the narrow-line region remains in the framework of
the host galaxy.
A few  candidates have been reported so far in the literature, but equally plausible  alternative interpretations for
these observations are still viable \citep[e.g.][]{robinson10,civano10,steinhardt12,markakis15,koss14,kalfountzou17}.

Recently, we found that the radio-loud QSO 3C~186 (z=1.068) displays all of the expected properties of a kicked active
SMBH \citep{pap3c186}.
The HST image taken with WFC3-IR at $\sim 1.4\mu$m clearly shows that the QSO does not reside at the center
  of the host galaxy, which appears to be a massive, relatively relaxed elliptical located at the center of a well studied X-ray
  cluster of galaxies
  \citep{siemiginowska05,siemiginowska10}.
  Despite the complexity of its spectrum, we measured significant velocity offsets between the UV broad (Ly$\alpha$, C~IV, C~III]
  and Mg~II) and narrow  lines.
   However, the available information did not allow us to definitely rule out
  other interpretations such as e.g. a chance superposition of a large elliptical with an under-massive QSO host galaxy and/or the
  presence of significant outflows in the QSO spectrum that may mimic the measured offsets. Prominent outflows have been observed
  in other QSOs, but they are typically seen in either the broad high ionization lines \citep[e.g.][]{shen16}
  or in narrow lines such as [OIII]5007 in the most powerful quasars known \citep[e.g.][]{vietri18}.
 Even if these alternative explanations
  seem to be unlikely for reasons that are
  extensively discussed in \citep{pap3c186}, it is extremely important to find independent
  confirmation of this GW recoiling BH candidate.

  In this work, we focus on one specific test that may shed further light on the structure of the emission
  line systems in 3C~186. We present the results of our integral field spectroscopy observations taken with the Keck telescope and
  OSIRIS. The main goal of these observations is to establish the spatial location of the narrow emission line features with respect to the location of the QSO. Are
  the narrow lines produced co-spatially with the QSO's continuum point source? Are lines of different widths/velocities produced in
  the same region? Can we spatially separate outflowing components and find evidence for quasar feedback in this source? Does the
  broad H$\beta$ line show the same offset as the UV permitted lines?

  In Sect.~\ref{observations} we describe the observations and the data reduction;
  in Sect.~\ref{results} we present the results,
  we discuss our findings in Sect.~\ref{discussion}, and in Sect.~\ref{conclusions} we draw conclusions.

Throughout the paper, the systemic redshift of the target is assumed to be z$_s$ = 1.0685, as derived in \citet{pap3c186} based on
both UV absorption lines and low-ionization narrow emission lines. We use the  following cosmological
parameters are used throughout the paper: H$_0$=69.6 km/s/Mpc, $\Omega_{\rm M} =0.286$, $\Omega_{\lambda} =0.714$.

\section{Observations and data reduction}
\label{observations}
The observations were performed on November 15, 2016 with OSIRIS \citep{osiris06}
and the Keck Laser Guide Star Adaptive Optics System. We utilized the Z$_{bb}$ filter which samples the
range of wavelengths between 999 and 1176nm. The scale used is 0.1''/pixel, and the corresponding
field of view is 1.6'' $\times$ 6.4''. The tip-tilt star correction was carried out using the quasar itself.
The observations were performed using a standard A-B-A-B sequence, each with an exposure time of 900s.
We collected a total of seven exposures on the target for a total of 6300s.
A standard star for telluric correction was also observed for 10 seconds.

The data reduction was performed using the OSIRIS data reduction pipeline (ODRP) v4.0.0 following the
procedures outlined in the ODRP Cookbook\footnote{\url{https://www2.keck.hawaii.edu/inst/osiris/drp_cookbook.html}}.
The steps include combining the sky exposures to make a sky image, reducing the telluric star data,
performing a basic reduction of the target and producing a data cube, correcting sky subtraction
to remove residual background, and correct for telluric absorption. For the reduction of the
telluric star we assumed a black body temperature of 10,000K.

\begin{figure}
  \plotone{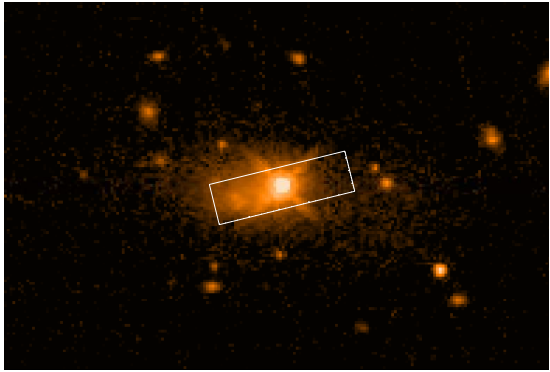}
  \caption{The OSIRIS field of view (1.6''$\times$ 6.4'',
    corresponding to $\sim 13$ kpc $\times \sim 52$ kpc, at the distance of the target) overlayed onto
    the HST/WFC3-IR F140W image of 3C~186  \citep[adapted from][] {pap3c186}.
     \label{osiris_fov}}
\end{figure}

In Fig.~\ref{osiris_fov} (left) we show the OSIRIS field of view overlayed onto the HST WFC3/IR image of the target
\citep{pap3c186}.

\begin{figure*}
  \gridline{\fig{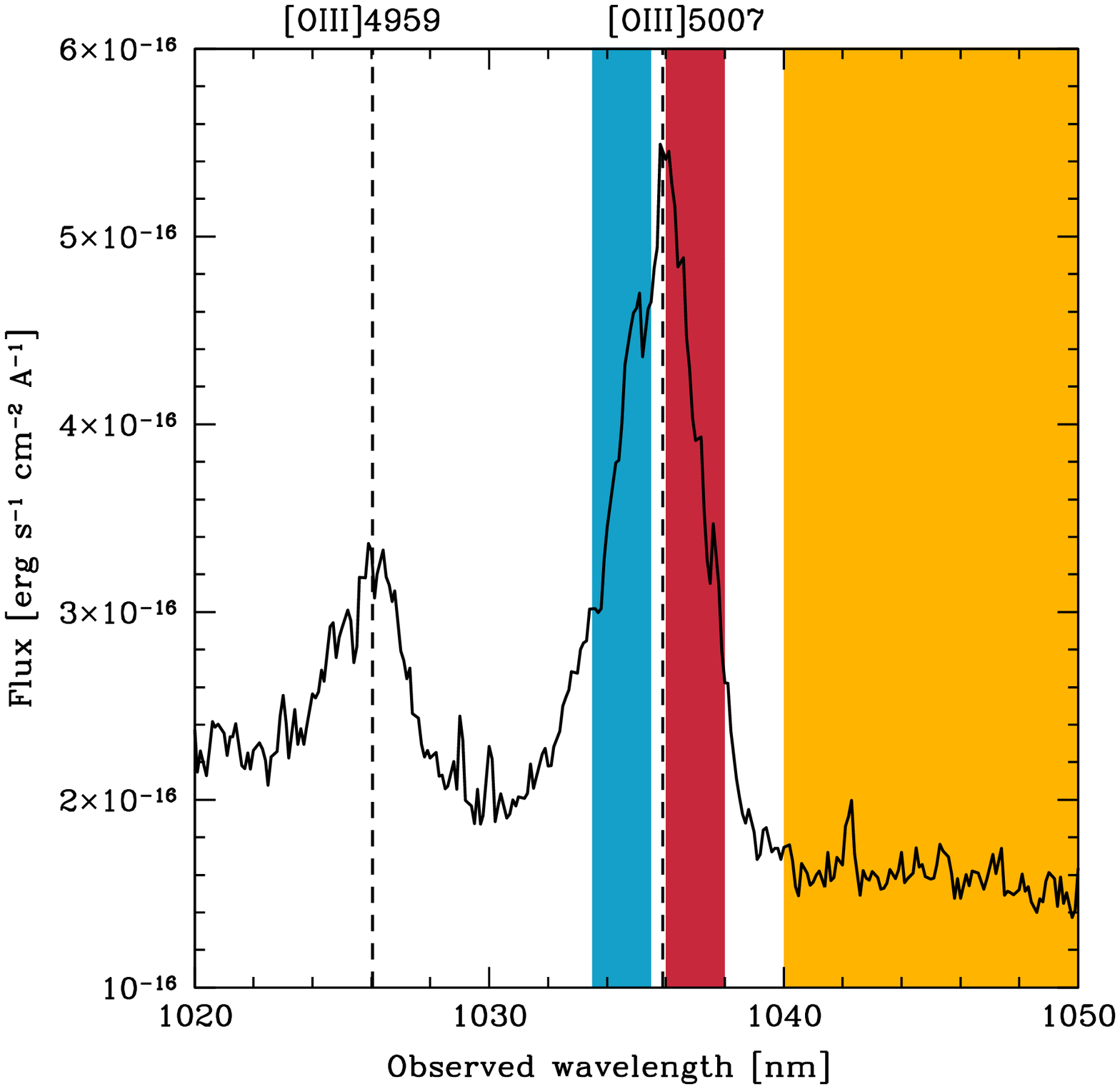}{0.5\textwidth}{}
    \fig{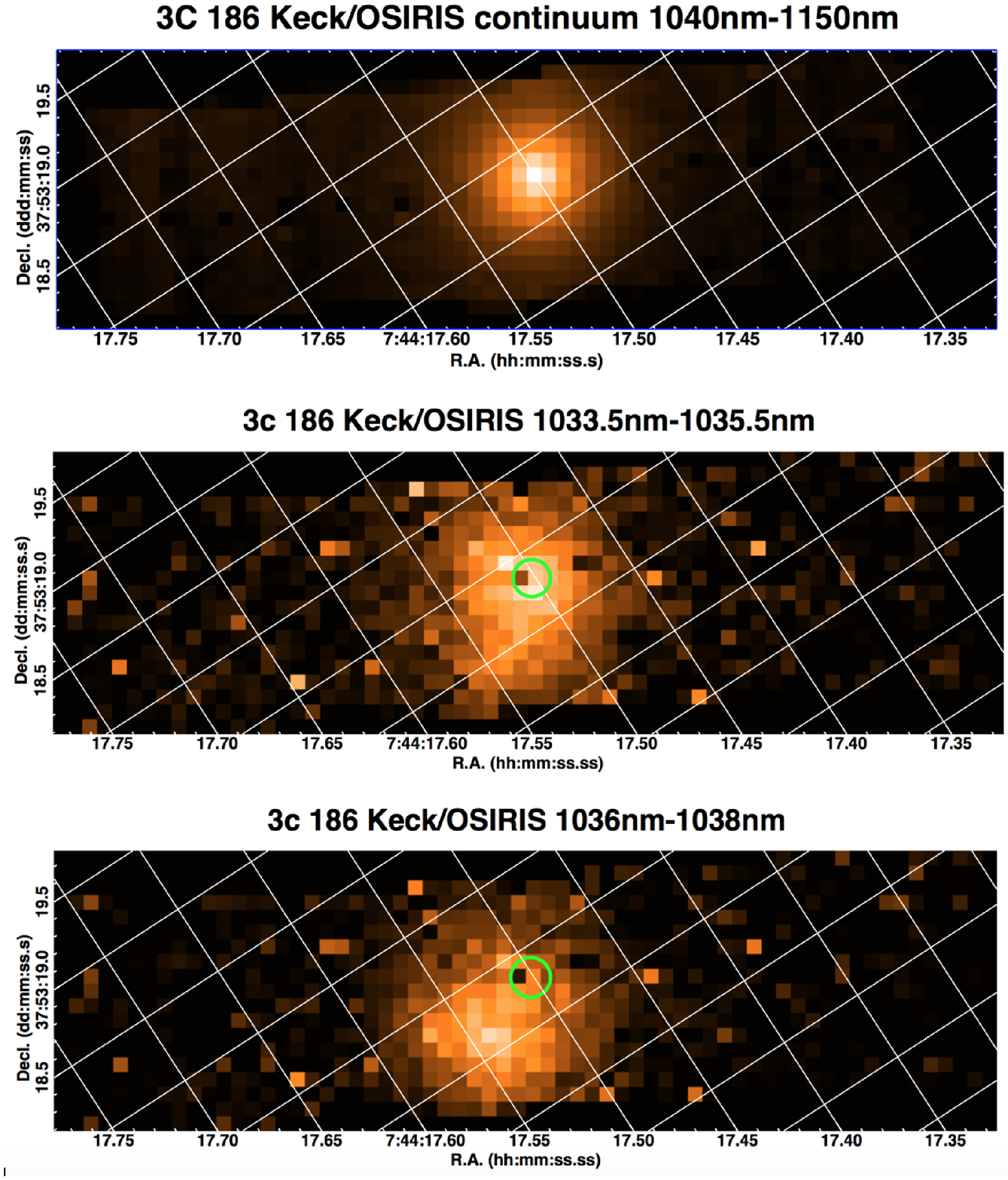}{0.45\textwidth}{}
  }
  \caption{{\bf Left:} Keck/OSIRIS spectrum of the region of the [OIII]4959,5007 doublet
    extracted from a region of r=6 spaxels (r=0.6'') centered
    on the centroid of the continuum source. The colored areas refer to the collapsed images shown
    in the right panel. Dashed lines indicate the redshifted wavelengths of the lines at the
    systemic redshift z$_s = 1.0685$. {\bf Right:} Keck/OSIRIS collapsed images of
    three different spectral regions. Continuum emission between 1040 and 1150nm
    (top, yellow shaded area in the left figure),
     continuum-subtracted blue side of
     the [OIII]5007 line between 1033.5nm  and 1035.5nm (center), and continuum-subtracted red side of the same
     line between 1036 and 1038nm (bottom). Green circles in the center and bottom panels indicate
    the centroid of the continuum source. \label{oiii}}
\end{figure*}
    
The absolute astrometry of the Keck data-cube was significantly off ($\sim 0.5$ arcmin) most likely
because of an error in the CRVAL keywords (i.e. the coordinates of the reference pixel). Therefore, we use
the collapsed continuum image to register the OSIRIS data on the WCS of the HST image. We make the
reasonable assumption that the point source seen in the HST data (i.e. the quasar)
is the continuum source in Keck/OSIRIS.

The measured FWHM of the continuum source in the spectral region between 1040 and 1150nm
is 0.55'', which we assume as the angular resolution of our dataset at that wavelength. In the following,
we will refer to this component simply as the QSO.

We performed a rough flux calibration by cross-calibrating the OSIRIS spectra with our
Palomar TripleSpec spectroscopic observations of the same target \citep{pap3c186}.
However, the various line components and the continuum emission
are not co-spatial, and the spectra extraction regions do not include the entire PSF. Therefore, we only
report flux densities in physical units where those represent a relatively accurate estimate of
the aperture-corrected values (e.g. in Fig.~\ref{oiii}, left panel).
Note that the aim of this work is not limited by the lack of an accurate flux calibration, since
we are mainly focused on the offset and the spatial location of each line component.

To register the data-cube onto the WCS of the HST observations and produce collapsed images we
utilize the {\it GAIA Starlink} software \citep{gaiastarlink}. To extract spectra at different locations
we use QFitsview \citep{qfitsview} with the {\it median} option.

\section{Results}
\label{results}
\subsection{Wavelength-dependent morphology}
The source presents a complex structure and its morphology varies depending on the
observing wavelength. The target is unresolved in the spectral region red-ward of the [OIII]5007 line,
which is dominated by the quasar continuum. No stellar emission from the host galaxy is detected across
the field of view.
The FWHM of the object in the spectral region
dominated by the broad H$\beta$ emission line between 999 and 1004nm is also consistent with
a point source (FWHM = 0.59'').

The detected emission lines in the wavelength range covered by these observations
are [OIII]5007,4959 and H$\beta$, superimposed to a strong power-law continuum. 
The most interesting spectral region is that of the [OIII]5007 line, which shows complex line
profile and morphology.

We first extract the spectrum from a relatively large region with radius r=6 spaxels (corresponding to
0.6'', i.e. approximately equal to the seeing). The center of the extraction region is fixed at the
spatial location of the quasar, i.e. on the centroid of the continuum emission as measured from the
collapsed image between 1040nm and 1150nm. The spectrum is shown in the left panel of Fig.~\ref{oiii},
limited to the region of the [OIII]4959,5007 doublet.
The [OIII]5007 line profile
can be decomposed into possibly two narrow components, and a blue wing. These line components
will be further discussed in the following (see Sect.~\ref{specfits}).
We derive three collapsed images from the interesting spectral regions
in which the S/N is sufficient to provide a
meaningful spatial analysis. We show the results in the right
panel of Fig.~\ref{oiii}: 
i) continuum emission red-ward of the [OIII]5007 line, top panel.
Part of this continuum spectral region is shown by the yellow area in the spectrum
reported in the left panel of the figure;
ii) continuum-subtracted blue side of the [OIII]5007 line, central panel (blue area in the left panel);
iii) continuum-subtracted red side of the [OIII]5007 line, bottom panel (red area in the left panel). The width
of each slice derived across the [OIII] line (i.e. regions ii and iii) corresponds to a velocity width of $\sim 600$ km s$^{-1}$.

The three spectral components correspond to three different regions of the target. We utilize SExtractor
within GAIA/Starlink for object detection. The FWHM of each source is also derived using the same
software.
As already pointed out above, the continuum source corresponds to the smallest observed structure and it is assumed to be spatially unresolved.
The blue side of the [OIII] line is emitted in a region
apparently centered on the quasar, but its
FWHM is $\sim$1.6'', i.e. significantly larger than the PSF. The red side of the
same line is also resolved, and most interestingly it is clearly off-nuclear.
Its brightest area is located $\sim$0.5'' South-East of the quasar.

We also attempted to derive a collapsed image for the bluer region of [OIII]5007 between 1031 and 1033nm,
to determine the spatial location and size of the blue wing of such a line. 
The derived continuum-subtracted source is only detected at a level of
$\lesssim 2.5 \sigma$ in that wavelength range, and thus
the spatial extension cannot be measured robustly with this method. In the following,
we derive more information on this component from the analysis of the nuclear and off-nuclear spectra.

It is worth noting that velocity and dispersion maps for these observations do not provide
  any useful information. This is likely because of both the complex nature of the spectrum, in which multiple
  lines are heavily blended, and  the low S/N.

\subsection{Nuclear and off-nuclear spectral fits}
\label{specfits}

\begin{figure*}
  \plottwo{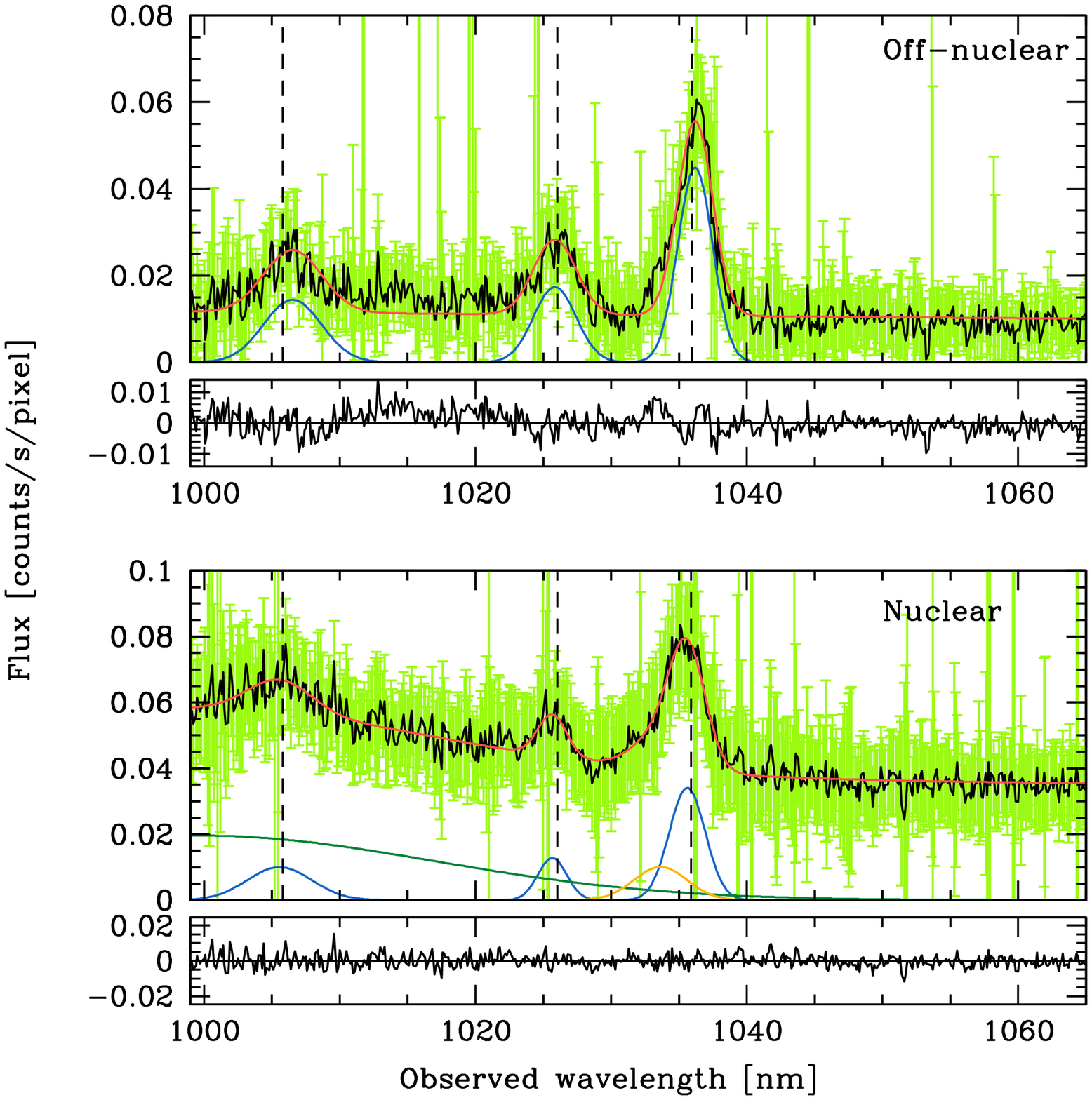}{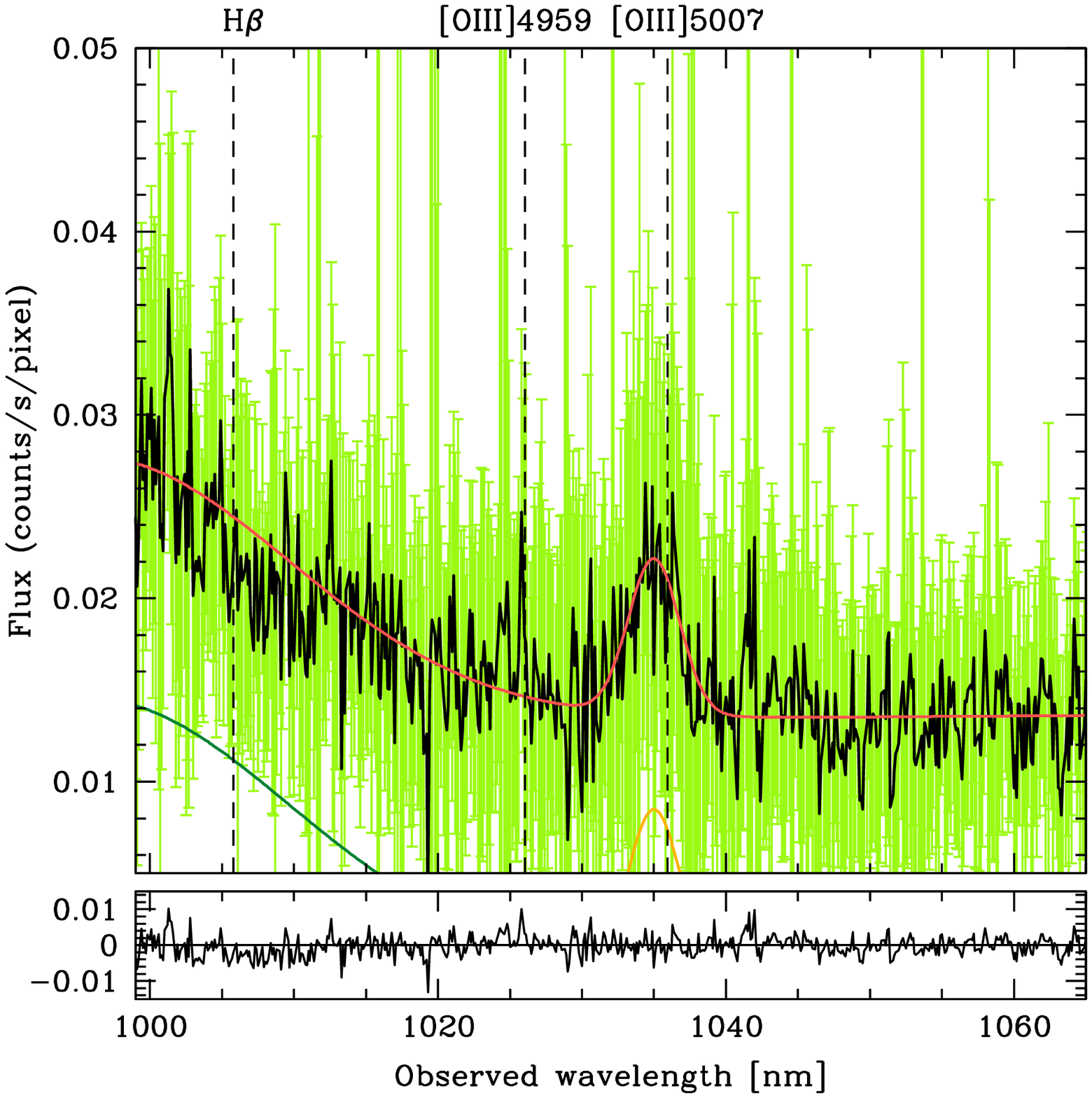}
  \caption{{\bf Left:} Nuclear and off-nuclear spectra are shown in the bottom and top panels, respectively.
    The nuclear spectrum is extracted from a region centered on the centroid of the continuum source. The off-nuclear spectrum is
    extracted from the region corresponding to the extended NLR located $\sim$ 0.5'' SE of the QSO (see Fig.~\ref{oiii}, bottom-right panel).
    {\bf Right:} Nuclear spectrum obtained by subtracting an annulus of width d=2 spaxels from the central r=2 spaxels centered on the QSO.
    Residuals after spectral model subtraction are shown below each spectrum. The range shown in the residuals box corresponds to the 1$\sigma$ error. \label{spec1}}
\end{figure*}

We fit spectra extracted at different positions using the {\it Specfit} tool in {\it IRAF}.
We utilize a power-law component and a  collection of Gaussian
profiles  to  fit each  line  of   interest.  The  parameters  are  then
successively  freed and  optimized through  a maximum of 100  iterations using a combination
of the {\it Simplex} and {\it Marquardt} minimization algorithms.  
The optimal parameters  for  each  line  are  determined  until  convergence  is
achieved. The initial guess on the fit parameters is made by  setting the
parameters to reasonable values. We then allow the parameters to vary until convergence is achieved,
without imposing any constraints on either the central wavelength or the FWHM of each line.

In Fig.~\ref{spec1} (left, bottom panel) we show the extracted spectrum of the
central region (i.e. r=1 pixel centered on the centroid of the continuum source, at
spaxel coordinates 11,32).
Errors (light green bars) are derived from the error array of each spaxel in
the FITS data-cube, and added in quadrature. By comparing
the errorbars with the dispersion of the datapoints we think that the errors derived from the datacube
are significantly overestimated.

The nuclear [OIII]5007 line profile clearly shows a blue wing, and the H$\beta$ line shows both a narrow
and a broad component. We fit the spectrum using five Gaussian components:
two for [OIII]5007,
one for [OIII]4959, and two for H$\beta$ (narrow and broad). Note that the [OIII]5007 and 4959 lines are expected to
show exactly the same components while in our model we only include one component for the fainter [OIII]4959 line,
for the sake of simplicity. A model using a single line is sufficient because of the the lower S/N in the spectral region covered by this line. The inclusion of any additional components fixed at the (rescaled) values derived for the [OIII]5007 line is a viable option but it does not result in any additional information.

The results of the best fit are shown in Table~\ref{tabfit}. 
The central wavelengths of the two components of the [OIII]5007 line are offset with respect to the
systemic redshift z$_s$ by -100 km s$^{-1}$ and -670 km s$^{-1}$. The former component is narrow (FWHM $\sim$ 900 km s$^{-1}$)
and the latter is slightly broader (FWHM $\sim$ 1300 km s$^{-1}$). The narrow component of H$\beta$ 
line also shows a small
blue-shift that is consistent with that of the narrow component of [OIII]5007.
The broad H$\beta$ line (FWHM $\sim$ 12000 km s$^{-1}$)
is only partially included in the wavelength range covered by OSIRIS and the Z$_{bb}$ filter.
The model clearly shows that the line is significantly blue-shifted (-1790$\pm$390 km s$^{-1}$).
The significance of such a blue-shift with respect to the systemic redshift z$_s$ is thus 4.6$\sigma$.

In Fig.~\ref{spec1} (left, top panel) we show the off-nuclear spectrum extracted from a circular region
of radius r=1 spaxels centered at the peak of the off-nuclear emission at coordinates (7,35).
In this case we only need three Gaussian lines to achieve a satisfactory fit.
The best fit shows that the most prominent line  (i.e. [OIII]5007) is slightly
red-shifted (+75$\pm$11 km s$^{-1}$). Note that the region of extraction of this off-nuclear spectrum
is still within the area covered by the PSF wings of the nuclear component. Therefore, the spectral
region between H$\beta$ and [OIII]4959 is most likely contaminated by the broad nuclear component
of the H$\beta$ line. Similarly, the continuum detected red-ward of the [OIII]5007 line is most
likely due to the PSF wings of the quasar continuum emission.

In the right panel of Fig.~\ref{spec1} we show the spectrum obtained by subtracting an annulus
of width d=2 spaxels just outside the nuclear spectrum extracted from a circular region of aperture r=2 spaxels,
centered on the brightest pixel of the continuum source.
The narrow components of all lines are no longer present, and only the broad H$\beta$
and the broad-ish [OIII]5007 (in addition to the nuclear continuum) are still visible.
On the one hand, this confirms that the broad component of H$\beta$ is produced in an
unresolved nuclear region. On the other hand, it
indicates that the same holds for the broad-ish component of [OIII].

We also fit the spectrum freezing the emission line wavelength of the broad H$\beta$ to the value
corresponding to the systemic redshift (1005.8nm). The model reproduces the data less accurately at the blue end of the spectral coverage,
and the $\chi^2$ is worse than for the best fit model, but
a $\chi^2$ difference test cannot completely rule out this scenario (P $<$ 0.22). On the contrary, for the nuclear spectrum shown in the
right panel of Fig.~\ref{spec1}, the same test shows that a broad H$\beta$ line at the systemic redshift is
completely inconsistent with the data. The $\chi^2$ difference test shows that the confidence level is extremely
high (P $< 5\times 10^{-5}$). 
While these results are clearly pointing towards a confirmation of an offset BLR, we stress that
we cannot draw definitive conclusions based solely on these data.
One major limitation of this dataset is that the H$\beta$ line is not entirely sampled by these observations, in
addition to the poor S/N in particular at the blue end of the spectral region covered by the $z_{bb}$ filter.

In order to investigate whether the fit is unique and if different initial conditions may affect the
best fit solution for the broad H$\beta$ component in particular, we perform the test described in the following.
We allow all parameters to be free to vary and we set the initial condition of the
central wavelength for the broad H$\beta$ line at
ten different values in the range $\lambda_c$= 997.0 -1006.0 nm.
We ran {\it Specfit} multiple times, using the {\it marquadt}
minimization method for all the values in the range reported above, until convergence is achieved. 
The result is that for all of the
initial conditions the solution converges to values in the range $\lambda_c$ = 997.4 - 999.7nm, in agreement
with the value reported in Table~\ref{tabfit}. The $\chi^2$ values are all indistinguishable and
the error are between 2.5nm (obtained for a best fit value of $\lambda_c$ = 999.7nm) and 0.7
(for $\lambda_c$ = 997.9nm). 
Therefore, we conclude that the results of the fit are relatively robust even if the noise in that spectral
region is high. Clearly, the lack of knowledge on the blue side of the line does not allow us to firmly conclude that
the best fit is unique.

We also checked that rebinning the dataset (in either the spatial or wavelength axes) does not produce more
accurate results because of the reduced resolution.

\begin{figure}
  \plotone{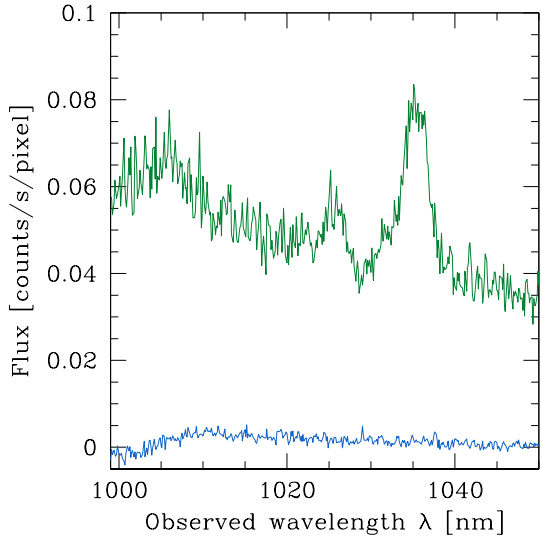}
  \caption{Average background spectrum extracted from four different empty regions of the detector, each with aperture
    radius r=2 spaxels (blue line), to investigate possible calibration issues at the blue end of the
    detector sensitivity band. For comparison, the green line shows the nuclear spectrum of 3C~186 (also
    shown in Fig.~\ref{spec1}). \label{background}}
\end{figure}

It is possible that small calibration issues may affect the shape of the spectrum at its blue end.
  In order to test such a hypothesis, we average the spectra of four empty regions of the field of view of the detector,
  each with an aperture radius r=2 spaxels.
  For wavelengths $< 1005$nm, i.e. at the very end of the bandpass, a small depression followed by a flux increase is observed
  (blue line in Fig.~\ref{background}).
  The amount of such an effect is ~0.005 c/s/pixel, peak-to-peak. This does not appear to be sufficient to significantly alter the
  shape of the extracted nuclear spectrum of the QSO. However, it is possible that the true line profile
  is slightly more peaked than reported in extracted nuclear spectrum (green line in Fig.~\ref{background}), and 
  it could be in fact similar to that derived for the annulus-subtracted nuclear spectrum shown in Fig.~\ref{spec1}, right panel.

\begin{deluxetable*}{l c c c c c c c c}
\tablecaption{Emission lines best fit parameters \label{tabfit}}
\tablehead{
  \colhead{Line} & \colhead{Observed wavelength} & \colhead{Err.} & \colhead{redshift}
  & \colhead{Err.}  & \colhead{Offset} & \colhead{Err.} & \colhead{FWHM} &	\colhead{Err.} \\
  \colhead{ } & \colhead{$\lambda$ [nm]}    & \colhead{} & \colhead{z} & \colhead{}
  & \colhead{[km s$^{-1}$]} & \colhead{ } & \colhead{[km s$^{-1}$]} \\
 }
\colnumbers
\startdata
Nuclear spectrum & & & & & & & & \\
\hline
H$\beta$ (narrow)        & 1005.57   &  0.35  & 1.0679 & 0.0007  & -82   & 104.0 & 1671  & 333   \\
H$\beta$ (broad)         & 999.86    &  1.35  & 1.0562 & 0.0028  & -1785 & 405.0 & 12067 & 1058  \\
{[OIII]}4959             & 1025.67   &  0.14  & 1.0678 & 0.0003  & -107  & 41.0 & 730   & 93    \\
{[OIII]}5007 (narrow)    & 1035.61   &  0.13  & 1.0678 & 0.0003  & -100  & 38.0 & 905   & 58 \\
{[OIII]}5007 (broad-ish) & 1033.64   &  1.01  & 1.0639 & 0.0020  & -670  & 293.0 & 1341  & 200 \\
\hline
\hline
Off-nuclear spectrum & & & & & & & & \\
\hline
H$\beta$ (narrow)       & 1006.52   & 0.18 & 1.0699 & 0.0004  & 200  & 54.0  & 1498 & 276 \\
{[OIII]}4959            & 1025.85   & 0.11 & 1.0681 & 0.0002  & -54  & 32.0  & 1078 &  90 \\
{[OIII]}5007 (narrow)   & 1036.20   & 0.04 & 1.0690 & 0.0001  &  71  & 12.0  & 872  &  33 \\
\hline
\hline
Annulus-subtracted nuclear spectrum & & & & & & & & \\
\hline
H$\beta$ (broad)         & 996.01   & 1.15  & 1.0483 & 0.0024  & -2933 & 346.0 & 9692 & 486 \\
{[OIII]}5007 (broad-ish) & 1034.99   & 0.21 & 1.0657 & 0.0004  & -279  & 60.0 & 1213 & 163 \\
\enddata
\end{deluxetable*}



\subsection{Energetics of the outflow}

Studies of the kinematics of the ionized gas around the central regions of active galaxies
often reveals the presence of significant velocity offsets that are interpreted as evidence for winds
\citep[e.g.][]{liu13,sun17,rupke17,ramosalmeida17}. This phenomenon constitutes one of the expected manifestations
of AGN feedback.
One of the main goals of
these observations is to study the presence of spatially displaced components of the narrow emission lines. As we noted
above, we discovered three different emitting regions associated with different spectral components of
the [OIII]5007 line. It is likely that these red and blue-shifted features are signatures of the presence of
winds powered by the active nucleus. It is interesting to compare the properties of such features with
other quasar-powered winds to determine whether 3C~186 is a typical object or it presents significant
peculiarities in its NLR structure, possibly because of its recoiling black hole.

We derive the maximum velocity v$_{max}$, mass rate $\dot{M}$,
and kinetic power L$_{kin}$ of the outflow using the formulae
reported in \citet{carniani15,bischetti17}. These authors studied samples of high-redshift (z$\sim 2.3- 3.5$)
hyper-luminous (L$_{bol} \sim 10^{47}-10^{48}$ erg s$^{-1}$) quasars and found relatively broad (FWHM $\sim$ 
1000-2000 km s$^{-1}$) components of the [OIII]5007 emission line. The maximum velocity of the outflow is defined as
v$_{max} = 2\sigma + |\Delta v| $, where $\sigma$ is the dispersion of the broad component of the [OIII] line,
and $\Delta$v is the velocity offset of that component with respect to the systemic redshift. For 3C~186,
v$_{max} = 1810$ km s$^{-1}$.
The mass outflow rate is 3300 M$_\odot$/yr, using formula (4) in \citet{bischetti17} and assuming the same parameters
as those authors except for a slightly smaller size of the emission line region (i.e. 4.8kpc instead of 7kpc)
because the higher spatial resolution of our Keck data allow us to set a slightly more stringent 
limit to the spatial scale.

The derived kinetic power is thus L$_{kin} \sim 3 \times 10^{45}$ erg/s. This corresponds to $\sim 4\%$ of the bolometric
luminosity for this object, which is L$_{bol} = 7.5 \times 10^{46}$ erg/s,
as estimated from the total [OIII] line luminosity \citep{pap3c186}.

Interestingly, \citet{bisogni17} found a inverse dependency between the EW of the [OIII]5007 line and the velocity offset
of the blue component of the same line in a large sample of SDSS QSO. Objects with large EWs display smaller velocity offsets.
For 3C~186, the EW as measured from the full aperture spectrum is $\sim 90$, while the offset is $-670$ km s$^{-1}$. This is about a
factor of six larger than the offset displayed by SDSS quasars with the same EW([OIII]). The same holds for the offset of
the main component of [OIII]. Our measured value of $-100$ km s$^{-1}$ is $\sim 5\times$ larger than observed in SDSS QSOs.
However, as discussed below, we stress that our target is significantly more powerful than the SDSS QSOs studied in \citet{bisogni17}.


\section{Discussion}
\label{discussion}
\begin{figure*}
  \plotone{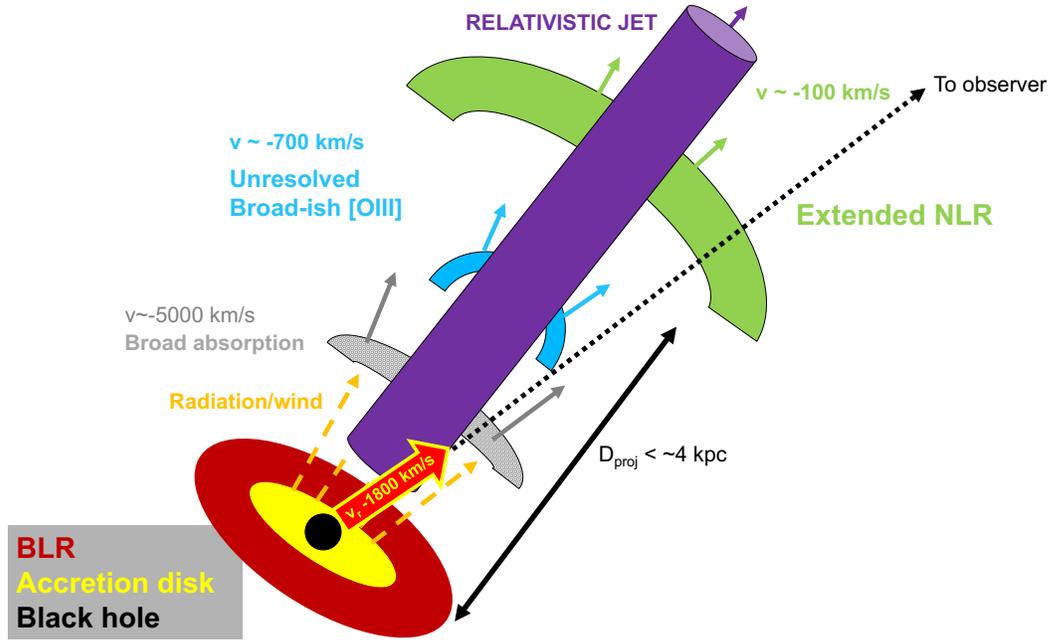}
  \caption{Schematic of the source as derived from the Keck/OSIRIS IFU data. Only the side of the AGN that faces the observer is reported.
    The green structure represents the slower (blue-shifted) outflowing component seen in the [OIII]5007 line. The blue feature is the $\sim 700$ km s$^{-1}$
    outflow, and the gray shaded area represents the fast ionized outflow that produces the broad absorption lines
    observed in the UV spectrum \citep{pap3c186}. The purple region is the relativistic jet. The black hole, the accretion disk and the BLR are
    shown in black, yellow and red, respectively. The radial velocity v$_r$ of the black hole and the BLR along the line-of-sight is also shown with
    a thick red-colored arrow \label{geometry}}
\end{figure*}

The Keck/OSIRIS IFU observations we present in this paper show three important properties of the target.
i) 3C~186 displays strong emission lines with multiple components; ii) the emission line region has
a complex morphology, similar to that observed in other powerful QSOs \citep[e.g.][]{liu13,carniani15}.
iii) The broad H$\beta$ line is blue-shifted with respect to the systemic redshift.

The three components of the [OIII]5007 emission line are produced in spatially distinct regions of the source. The most prominent
component is red-shifted by $\sim$ 70 km s$^{-1}$, its FWHM is $\sim$ 900 km s$^{-1}$, and it is produced in a spatially
resolved off-nuclear region, at a distance of about 0.5'', corresponding to a projected distance of $\sim$ 4kpc.
This feature is approximately located along the direction of the radio jet, which is oriented along a NW-SE direction.
In low-frequency VLBI maps \citep{nan92}, the jet is  one sided
and the SE side of the jet (i.e. the counter-jet) is not detected. Therefore any spatial association between the radio jet and
the extended NLR cannot be firmly determined. However, it seems plausible that the red-shifted component of the [OIII] line could
be produced in a relatively slow outflow possibly caused by jet-induced feedback, along the counter-jet
\citep[e.g.][]{odea98,odea03}.

There are two blue-shifted components of the [OIII]5007 line. The former is narrow (FWHM $\sim$ 900 km s$^{-1}$),
its offset is $\sim$ -100 km s$^{-1}$, and the emitting region is resolved at the angular resolution of our IFU data. It is natural
to identify such a component with the blue-shifted counterpart of the narrow, red-shifted feature of the same line. Its centroid
is approximately aligned with the the center of the QSO, therefore  it must be produced in a region that is
close, in projection,  to the accretion disk. We stress that the angular resolution of our data corresponds to a few kpc,
therefore we can only set an upper limit at a level that is allowed by our data.

The latter component is significantly broader (FWHM $\sim$ 1300 km s$^{-1}$), its emitting region is
likely very compact (i.e. unresolved in our data) and its location
coincides with the QSO continuum emission. The (blue-)shift displayed by such a component
is $\sim -700$ km s$^{-1}$, therefore
we conclude that this is a relatively fast wind produced even closer to the accretion disk as compared to the previously
discussed features, but clearly still well outside of the BLR.

The second major result of these observations is the evidence in support of an extremely large blue-shift of the emission lines
emitted within the BLR. The H$\beta$ line is only partially included in our data, because at the redshift
of the source its central wavelength is expected to be very close to the blue end of the sensitivity pass-band of the instrument.
Despite the low S/N, we showed that
the offset is significant, at a level of $\sim 4.6\sigma$. The best-fit value of the offset (v=$-1790 \pm 400$ km s$^{-1}$) is consistent with the
results of \citet{pap3c186} based on the analysis of broad UV lines (Ly$\alpha$, C~IV and MgII, v=$-2140 \pm 390$ km s$^{-1}$), within the margin of error.
Note that those UV lines display a concave shape on the
blue side, which was explained as due to a broad absorption feature.
We cannot confirm such a feature to be present in the H$\beta$ line with these data, since we are only sampling the red side of
that line. Therefore, the results presented here are derived under the assumption of a symmetric Gaussian shape for each line.

Even if the statistical evidence for a $\sim -1800$ km s$^{-1}$ velocity offset is
quite robust, we need to be extremely careful in drawing firm conclusions solely based on the results presented here.
In fact, as we pointed out above, the peak of the line is located at the very end of the spectral region covered by
our observations, and that might significantly affect the results because of the rather poor S/N
in that region.

The fact that both the broad H$\beta$ line and the broad-ish [OIII] line are produced in an unresolved region of the AGN, co-spatial
with the QSO continuum (at the resolution of our data) is also indicated by the nuclear spectrum shown in Fig.~3 (right panel),
obtained after subtracting off the emission from an annulus of d=2 spaxels just outside the nuclear aperture. It is remarkable
that the only features that are left after that operation are the pure nuclear components, i.e. the QSO continuum, the broad line and
the fast  outflowing component of the [OIII] line.

The interpretation of our data is consistent with the picture envisaged by \citet{pap3c186}.
The BLR, which is attached to the black hole
and its accretion disk, is moving at high velocity with respect to the gas at rest in the host galaxy of this QSO.
This can be modeled as the result of a gravitational wave recoil kick, following the merger of two high-spin 
supermassive black holes of comparable mass \citep{lousto17}. 

The effect of AGN feedback is evident from both the morphology of the emission line region and the spectral complexity of the spectral lines. 
The object seems to agree with other samples of similar sources, the only peculiarity being the rather large EW of
the [OIII] line with respect to the velocity of the outflows, as compared to what is observed in a large sample of SDSS QSOs \citep{bisogni17}. However, 3C~186 is $\sim$ 1 dex more powerful than the brightest SDSS QSO considered in \citet{bisogni17},
in terms of AGN bolometric luminosity.
Therefore, a comparison with a sample of quasars with similar power \citep[e.g. such as that considered by][]{bischetti17}
is likely more appropriate to establish any peculiarities of our source. However, the fact that the general properties of the observed
outflows match those of the general QSO population implies that radiation pressure onto the ISM of the host is likely producing the winds.
Mechanical feedback seems ruled out because, in such a scenario, the added velocity of the kicked AGN  would imply a faster wind with respect
to non-recoiling AGNs of similar power.

In Fig.~\ref{geometry} we show a schematic representing a possible geometry of the QSO and the structure of the NLR. The green feature
is the slower [OIII] component (the extended NLR) that lies on the side of the jet that points towards the observer.
A similar feature (not shown in the figure) is also present on the
opposite side of the accretion disk with respect to the observer (i.e. the mildly red-shifted component of the [OIII] line).
Such a feature is likely located at a significantly larger distance from the BH, to account for the observed spatial offset.
The blue feature corresponds to the broad-ish [OIII] component, which is most likely
located within a few kpc from the BH. The gray shaded region represents the fast ionized outflow that produces the broad absorption lines
observed in the UV spectrum \citep{pap3c186}. The purple cylinder represents the direction of propagation of the radio jet. The dashed yellow
arrows represent radiation pressure exerted onto the host galaxy ISM, which is most likely producing the observed winds. At larger distances, it is
possible that the radio jet also contributes to this feedback process.

\section{Conclusions}
\label{conclusions}
We presented Keck/OSIRIS IFU data of the radio-loud QSO 3C~186 aimed at studying both the structure and kinematics of the narrow line region
of the source. HST observations first published in \citet{pap3c186} showed that the QSO is offset with respect to the center of the host galaxy
by $\sim 11$kpc. Rest-frame UV spectra presented in that paper also show a significant ($\sim 2100$ km s$^{-1}$)
velocity offset between the broad and narrow lines.
The observations were interpreted as the result of a gravitational wave kick resulting from a merger of two supermassive black holes
that happened $\sim 5\times 10^6$ years ago. Recoiling black holes originated by GW kicks are extremely important objects because of
their bearings on our knowledge of how massive black holes interact with each other, possibly merge and grow in size. Finding confirmed GW kicked BHs is also
crucially important to rule out the so-called final parsec problem which might prevent SMBHs from merging after galaxy merger events,
thus limiting the possibility that GWs emitted by such a phenomenon can be detected with pulsar timing array experiments and by future
space missions such as LISA. These Keck observations do not constitute a definitive proof that the interpretation as a GW recoiling
black hole for this object is
correct.
However, these data appear to support such a scenario. In addition, the effects of AGN feedback onto the galaxy ISM are
clearly visible.

We showed that there are three different spectral components, each associated with a spatially distinct region of the emission line region.
The analysis of the profile of the [OIII]5007 line shows that a relatively slow (v $\sim$ 100 km s$^{-1}$) outflow is present
on large scales ($>$ a few kpc from the BH). This outflow is both blue-shifted and red-shifted, and each component is associated with a
spatially distinct region, roughly aligned with the direction of the relativistic jet.
A faster, blue-shifted (v $\sim$ -700 km s$^{-1}$) outflowing component which is co-spatial with the QSO center is also seen. The properties
of the outflows observed in this QSO are consistent with those seen in other QSO samples, although some peculiarities are present.
Based on the observed properties, we conclude that radiation pressure onto the ISM is the most likely origin for this feedback process.

The broad line region, which is sampled in our observations by the H$\beta$ line, is significantly blue-shifted with respect to the narrow
emission lines. The H$\beta$ line is not fully sampled by our data, since its central wavelength apparently lies at the blue end of the
filter pass-band. However, the best fit model shows a statistically significant velocity offset of $\sim 1800$ km s$^{-1}$,
i.e. consistent with the findings of \citet{pap3c186} based on permitted UV lines.

Spectroscopic observations sampling the full H$\beta$ broad emission line will be key to definitely confirm the GW recoiling BH scenario. In order to
provide final evidence in favor (or against) this interpretation, James Webb Space Telescope IFU data should be obtained using the NIRSPEC IFU
instrument to fully sample the H$\beta$ spectral region and achieve a significantly higher spatial resolution (0.1'' vs. 0.5'' provided by the
observations presented here). Deep imaging with HST and JWST information will be extremely
important to further constrain the absence of a second (under-massive) galaxy around the QSO, as discuss at length in \citet{pap3c186}.
Finally, with JWST it will also be possible to detect spectral features from stars in the host galaxy, thus providing an independent
measure of the redshift of the host.

\acknowledgments

The authors wish to thank the anonymous referee for insightful comments that helped to
significantly improve the paper. MC thanks Marshall Perrin for useful hints on the data reduction.
The data presented herein were obtained at the W.M. Keck Observatory, which is operated as a scientific
partnership among the California Institute of Technology, the University of California and the National
Aeronautics and Space Administration. The Observatory was made possible by the generous financial support
of the W.M. Keck Foundation. The authors wish to recognize and acknowledge the very significant
cultural role and reverence that the summit of Maunakea has always had within the indigenous Hawaiian
community. We are most fortunate to have the opportunity to conduct observations from this mountain. 
Some of the data presented here are based on observations made with the NASA/ESA Hubble Space Telescope,
obtained from the data archive at the Space Telescope Science Institute. STScI is operated by the
Association of Universities for Research in Astronomy, Inc. under NASA contract NAS 5-26555.




\end{document}